\documentclass[prb,aps,twocolumn,superscriptaddress,showpacs]{revtex4}

\usepackage{amsmath}
\usepackage{amssymb}
\usepackage{graphicx}
\usepackage{subfigure}
\usepackage{bm}% bold math

\usepackage{color}
\usepackage{ulem}
\begin{document}

\bibliographystyle{apsrev}

\title{Asymmetric long Josephson junction acting as a ratchet for a quantum  field}

\author{A.O. Sboychakov}
\affiliation{Advanced Science Institute, the Institute of Physical
and Chemical Research (RIKEN), Wako-shi, Saitama, 351-0198, Japan}
\affiliation{Institute for Theoretical and Applied Electrodynamics
Russian Academy of Sciences, 125412 Moscow, Russia}

\author{Sergey Savel'ev}
\affiliation{Advanced Science Institute, the Institute of Physical
and Chemical Research (RIKEN), Wako-shi, Saitama, 351-0198, Japan}
\affiliation{Department of Physics, Loughborough University, Loughborough LE11 3TU, UK}

\author{A.L. Rakhmanov}
\affiliation{Advanced Science Institute, the Institute of Physical
and Chemical Research (RIKEN), Wako-shi, Saitama, 351-0198, Japan}
\affiliation{Institute for Theoretical and Applied Electrodynamics
Russian Academy of Sciences, 125412 Moscow, Russia}

\author{Franco Nori}
\affiliation{Advanced Science Institute, the Institute of Physical
and Chemical Research (RIKEN), Wako-shi, Saitama, 351-0198, Japan}
\affiliation{Department of Physics, the University of Michigan, Ann Arbor, MI 48109-1040, USA}


\begin{abstract}
We study the escape rate of flux quanta in a long Josephson junction having an asymmetric spatial inhomogeneous critical current density. We show that such a junction can behave as a quantum ratchet when it is driven by an ac current in the presence of a magnetic field. The rectification
gives rise to an onset of the dc voltage $V_{dc}$ across the junction. The usual approach of particle-like tunneling cannot describe this rectification, and a quantum field theory description is required. We also show that under definite conditions the rectification direction and, consequently $V_{dc}$, can change its sign when varying the temperature $T$ near the crossover temperature $T^{*}$ between the quantum and classical regimes.
\end{abstract}
%\pacs{74.50.+r, 85.25.Cp, 12.20.-m, 32.80.-t}

\date{\today}

\maketitle

\textit{Introduction}.--- Solid state devices with asymmetric periodic potentials (acting as ratchets) are attracting considerable interest~\cite{HMarcNori}. Ratchets can produce a direct current when a time-dependent force (deterministic or random) with zero mean is applied to it. This effect occurs for both: thermal hopping and quantum tunneling. Moreover, for some special shapes of the ratchet potential, the current can change its sign for decreasing temperatures, when quantum tunneling becomes dominant over thermal-activated hopping~\cite{HanggiPRL97}. There are many different realizations of ratchets, both in nature and in artificial nano-devices, such as cold atoms, colloidal magnetic particles, single-molecule optomechanical devices, fluxons in superconductors, and many other systems (for reviews, see, e.g., Ref.~\onlinecite{HMarcNori}). All these systems are effectively described by a quantum or classical particle moving in a periodic asymmetric potential.

Here we propose a completely new ratchet system, which is described by a quantum field. Namely, we consider a long Josephson junction (JJ) (the junction's length $D$ is comparable with the Josephson penetration depth $\lambda_J$) with a spatially modulated critical current density $i_c(x)$ driven by an adiabatically slow ac current $J=J_0\cos\Omega t$, where the amplitude $J_0$ is smaller than the junction's critical current $J_c$. The gauge-invariant phase difference $\varphi(x)$ plays the role of the field variable. Although $J$ never exceeds $J_c$, the flowing current can give rise to sudden changes in the phase difference, both due to the thermal hopping of the flux quanta through the potential barrier and quantum tunneling. The probability per unit time (escape rate) $\Gamma(J)$ of these transitions attains maxima $\Gamma_{\pm}$ at $J=\pm J_0$, and one can expect that $\Gamma_{+}=\Gamma_{-}$. However, the application of an external dc magnetic field $H$ to the junction having current inhomogeneity  asymmetric with respect to the $x$ direction gives rise to the asymmetry in tunneling probability, $\Gamma_{+}\neq\Gamma_{-}$. This leads to the appearance of a dc voltage $V_{dc}$ across the junction, since $V_{dc}=(\hbar/2e)\langle\frac{\partial\varphi}{\partial t}\rangle\propto(\Gamma_{+}-\Gamma_{-})$. We calculate the escape rate $\Gamma$ using the well-known method of imaginary-time trajectories at finite temperature~\cite{ColeICaldeira}. According to this approach, $\Gamma$ can be written as $\Gamma=A\exp(-B)$, where the prefactor $A$ is of the order of the Josephson plasma frequency $\omega_p$, and $B=S_{\beta}/\hbar$. Here, $S_{\beta}$ is the action of the system calculated for a periodic imaginary-time trajectory, $\varphi(\tau+\hbar\beta)=\varphi(\tau)$, with a period $\tau_0=\hbar\beta$, $\beta=1/T$. The effect of quantum tunneling in JJs and stacks of intrinsic JJs in high-$T_c$ superconductors is now studied intensively, experimentally~\cite{ExpMQT} and theoretically~\cite{TheoMQT}, both due to its fundamental interest and the possibility to use these systems in future applications. A ratchet based on a SQUID consisting of two equal JJs in series, coupled in parallel to a third junction, was proposed in Ref.~\onlinecite{HangiiSQUID}. However, the particle-like approach used in most theoretical considerations of quantum tunneling in JJs is not appropriate here, and a field-theoretical description must be used.

We stress that the system under consideration does {\it not} correspond to a ratchet in the usual sense since the junction's potential is the {\it periodic} sine-Gordon potential, $U(\varphi)\propto-\cos\varphi$, which is {\it symmetric} in $\varphi$. The difference between $\Gamma_{+}$ and $\Gamma_{-}$, and, consequently, a voltage rectification occurs here due to both: the ``parametric'' dependence of the potential $U$ on $x$, and the spatial dependence of the field $\varphi(x)$. Thus, the coordinate dependence of the phase difference is crucial for the rectification, even for relatively short junctions, when $D<\lambda_J$. Here we exploit the approach developed in Ref.~\onlinecite{weAll}, which we generalize here to finite temperatures. This now allows us to calculate, in the same manner, the escape rate $\Gamma$ both in the quantum and classical regimes. If the junction width $D$ is large, then the exponent $B$ of the escape rate $\Gamma$ turns out also to be large. However it should be not too large, when $\Gamma$ becomes an experimentally non-observable quantity. This last condition means that the current amplitude $J_0$ should be close to the critical current~\footnote{Strictly speaking, there are two critical currents (in magnetic field), $J_c^{\pm}$, corresponding to currents flowing in two opposite directions. The current $J_c=\text{min}(J_c^{+},J_c^{-})$.}. We focus on the exponent $B$ because changes in $B$ (not in the prefactor $A$) describes the main change in $\Gamma$ when $B$ is large. In addition, here we do not consider dissipation in the system when calculating $B$. The effect of dissipation does not change qualitatively the results obtained here.

\begin{figure}
\begin{center}
\includegraphics*[width=0.47\textwidth]{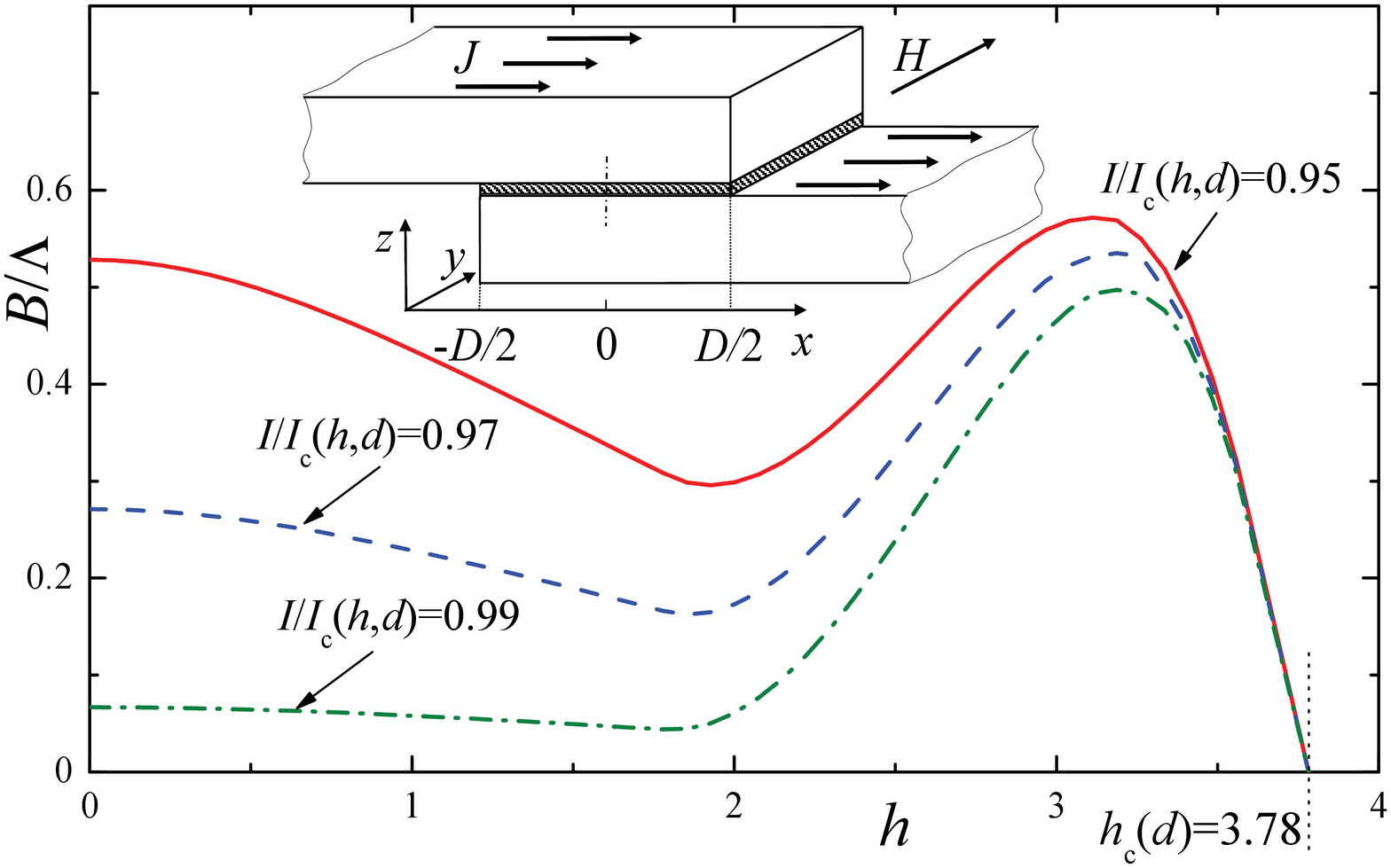}
\end{center}
\caption{\label{FigB0H} (Color online) The magnetic field dependence of the escape rate's exponent $B$ for fixed current ratios $I/I_c(h,d)$ in the uniform junction [$\gamma(x)=0$], calculated according to Eq.~\eqref{B}; $d=1.8$ and $T=0$. The inset shows the schematic geometry of the Josephson.}
\end{figure}

\textit{Escape rate}.--- We consider a Josephson junction in the {\it inline} geometry shown in the inset of Fig.~\ref{FigB0H}. Two superconducting bars overlap a length $D$ in the $x$-direction. The external magnetic field $H$ is applied in the $y$-direction. Let us first consider the junction biased by a {\it dc} current $J$. The generalization to the adiabatically varying ac current $J\cos\Omega t$ is evident, and the conditions of adiabaticity will be given below.  The spatial inhomogeneity of the critical current density $i_c(x)$ in the $x$-direction can be realized, e.g., by a spatial variation of the thickness of the insulating layer between superconductors, or slightly changing the shape of the junction itself. We assume that $J_c$ is homogeneous in the $y$-direction. The system under study can be described by an effective 1D Lagrangian for the phase difference $\varphi(\tau,x)$. The quasi-classical imaginary-time action of this system is
\begin{eqnarray}\label{S0}
{\cal S}_{\beta}[\varphi]&=&\frac{E_J}{\omega_{p}}\int_0^{\hbar\omega_p\beta}\!\!\!\!\!\!\!\!\!d\tau\left[{\cal L}[\varphi]+{\cal L}_{\Sigma}[\varphi]\right],\nonumber\\
{\cal L}[\varphi]&=&\int_{-d/2}^{d/2}\!\!\!\!dx\!\left[\frac12\left(\frac{\partial\varphi}{\partial\tau}\right)^2\!\!\!\!+%
\frac12\left(\frac{\partial\varphi}{\partial x}\right)^2\!\!\!\!-g(x)\cos\varphi\right],\nonumber\\
{\cal L}_{\Sigma}[\varphi]&=&\varphi\Big(\tau,-\frac{d}{2}\Big)\left[h-\frac{I}{2}\right]- \varphi\Big(\tau,\frac{d}{2}\Big)\left[h+\frac{I}{2}\right].
\end{eqnarray}
In Eqs.~\eqref{S0}, the $x$-coordinate is normalized by $\lambda_J$, $\tau$ by $1/\omega_{p}$, $d=D/\lambda_J$, $\beta=1/T$, $E_J=\hbar\overline{i_c}\lambda_JL/(2e)$ is the Josephson energy ($L$ is the junction's length in the $y$-direction), the bar above $i_c$ means spatial averaging, and
\begin{equation}
g(x)=\frac{i_c(x)}{\overline{i_c}}\equiv1+\gamma(x)\,,\;\;\;\overline{\gamma(x)}=0\,.
\end{equation}
The phase $\varphi(\tau,x)$ satisfies the equation of motion
\begin{equation}\label{EqPhi}
\frac{\partial^2\varphi}{\partial\tau^2}+\frac{\partial^2\varphi}{\partial x^2}-g(x)\sin\varphi=0\,,
\end{equation}
with the periodicity condition $\varphi(\tau+\hbar\omega_p\beta,x)=\varphi(\tau,x)$. The boundary conditions for the phase difference $\varphi(\tau,x)$ are defined by the ${\cal L}_{\Sigma}$ term in the action~\eqref{S0}: $\partial\varphi/\partial x|_{x=\pm d/2}=\pm I/2+h,$ where $I$ and $h$ are the dimensionless current and external magnetic field, respectively: $I=J/(\overline{i_c}\lambda_JL)$, and $h=cH/(4\pi\overline{i_c}\lambda_J)$.

In general, it is hard to find a periodic solution to the nonlinear Eq.~\eqref{EqPhi}, even numerically. However, as mentioned above, we only need to find a solution when the current $I\approx I_c$. In this case, we can use the approach developed in Ref.~\onlinecite{weAll}. Namely, we seek a solution of the form $\varphi(\tau,x)=\varphi_0(x)+\psi(\tau,x)$, where $\varphi_0(x)$ is the steady-state solution of Eq.~\eqref{EqPhi} corresponding to an energy minimum. Since $I$ is close to the critical current, the energy barrier between the neighboring energy minima, $\varphi_0(x)$ and $\varphi_0(x)\pm2\pi$, is small, $|\psi(x,\tau)|\ll\varphi_0(x)$, and we can expand the action~\eqref{S0} in powers of $\psi$ up to $\psi^3$. Then, we represent $\psi(\tau,x)$ in the form of a series
\begin{equation}\label{expPsi}
\psi(\tau,x)=\frac{3\mu_0}{U_{000}}\sum_{n=0}^{\infty}\alpha_n(\sqrt{\mu_0}\tau)\psi_n(x)\,,
\end{equation}
where $\mu_n$ and $\psi_n(x)$ are the eigenvalues and orthogonal eigenfunctions of the operator~\footnote{In real time, $\psi_n$ correspond to linear standing-waves with frequencies $\omega_n=\sqrt{\mu_n}\omega_p$ excited over the solution $\varphi_0(x)$.}
\begin{equation}
%\hat{{\cal D}}=-\frac{\partial^2}{\partial x^2}+g(x)\cos\varphi_0(x)\,,\,\,\,\,\hat{{\cal D}}\psi_n=\mu_n\psi_n\,,
\hat{{\cal D}}=-\partial^2/\partial x^2+g(x)\cos\varphi_0(x)\,,\,\,\,\,\hat{{\cal D}}\psi_n=\mu_n\psi_n\,,
\end{equation}
and
\begin{equation}
U_{nmk}=\int_{-d/2}^{d/2}\!\!\!\!dx\,g(x)\sin\varphi_0(x)\psi_n(x)\psi_m(x)\psi_k(x)\,.
\end{equation}

Expanding the equation of motion~\eqref{EqPhi} in powers of $\psi$, up to $\psi^2$, with $\psi$ in the form~\eqref{expPsi}, multiplying it by $\psi_n(x)$, and integrating over $x$, we obtain a system of ordinary differential equations for the collective coordinates $\alpha_n(\eta)$ of the field $\psi$, where we introduce the imaginary-time variable $\eta=\sqrt{\mu_0}\tau$. It can be shown that due to the proximity of $I$ to $I_c$, two conditions for the eigenvalues $\mu_n$ are possible: either $\mu_0\ll\mu_n$, $n>0$ for relatively short junctions ($d\lesssim1$) {\it or} large fields ($h\gtrsim1$), or $\mu_0\sim\mu_1\ll\mu_n$, $n>1$ for long junctions ($d\gtrsim1$) {\it and} small fields ($h\ll1$). Due to these inequalities, one can neglect all the equations in the system of equations for $\alpha_n$, except the first two (for details, see Ref.~\onlinecite{weAll}). Thus, the system of equations for $\alpha_0$ and $\alpha_1$ takes the form
\begin{equation}\label{EqAlpha}
\ddot{\alpha}_0+\frac{\partial V}{\partial\alpha_0}=0\,,\,\,\,\,%
\ddot{\alpha}_1+\frac{\partial V}{\partial\alpha_1}=0,\,
\end{equation}
where the dot means derivative over $\eta$, and the potential $V(\alpha_0,\alpha_1)$ can be written as
\begin{equation}
%V\!=\!\frac12\!\!\left[\!\alpha_0^3+v_{11}\alpha_1^3+3v_{00}\alpha_0^2\alpha_1+3v_{01}\alpha_0\alpha_1^2- \alpha_0^2-\frac{\mu_1\alpha_1^2}{\mu_0}\!\right]\!\!,
V=\frac12\sum_{i=0,1}\left[\frac{U_{iii}\,\alpha_i^3+3U_{01i}\,\alpha_0\alpha_1\alpha_i}{U_{000}}%
-\frac{\mu_i\alpha_i^2}{\mu_0}\right].
\end{equation}
We should find a periodic solution of the system~\eqref{EqAlpha} with period $\eta_0=\sqrt{\mu_0}\hbar\omega_p/T$ (``bounce'' solutions). The exponent $B=S_{\beta}/\hbar$ of the escape rate $\Gamma=A\exp(-B)$ can be expressed through the functions $\alpha_i(\eta)$, $i=0,1$ as:
\begin{equation}\label{B}
B=\frac{9\Lambda\mu_0^{5/2}}{(U_{000})^2}\int_0^{\eta_0}\!\!\!\!\!d\eta\left[\!\sum_{i}\frac{\dot{\alpha}_i^2}{2}-V(\alpha_0,\alpha_1)\!\right],\, \Lambda=\frac{E_J}{\hbar\omega_p}.
\end{equation}

Thus, we reduce the field-theory problem to the problem of one particle moving in the potential $V(\alpha_0,\alpha_1)$ in {\it two
dimensions}, where the $\alpha_i$'s play the role of the particle's generalized coordinates. Let us first consider the case of $T\to0$, when  quantum tunneling prevails. Our analysis shows that when $d>d_c(I,h)\sim4$, there exist three bounce solutions to the system~\eqref{EqAlpha}: $\alpha_i^{(s)}(\eta)$, $s=0,\pm1$. In analogy to the case $\gamma(x)=h=0$ considered in Ref.~\onlinecite{weAll}, one can say that the solution $\alpha_i^{(+1)}(\eta)$ ($\alpha_i^{(-1)}(\eta)$) corresponds to the formation of a vortex (antivortex) nucleus at the left (right) junction edge, while the solution $\alpha_i^{(0)}(\eta)$ describes the tunneling of $\varphi$ as a whole (for details, see Ref.~\onlinecite{weAll}). When $\gamma(x)=h=0$, the solutions have the following symmetric properties: $\alpha_1^{(0)}(\eta)=0$, $\alpha_1^{(-1)}(\eta)=-\alpha_1^{(+1)}(\eta)$. Thus we have three channels of tunneling, with probabilities $\Gamma^{(s)}\propto\exp(-B^{(s)})$, with $B^{(-1)}=B^{(+1)}$, and the total probability becomes $\Gamma=\sum_s\Gamma^{(s)}$. The applied magnetic field breaks the vortex-antivortex symmetry [$B^{(-1)}=B^{(+1)}$] making one of these channels more favorable. However, the total escape rate $\Gamma$ is still symmetric with respect to the direction of the dc current if $\gamma(x)=0$. For spatially inhomogeneous junctions with $\gamma(x)\neq0$, we have $\Gamma(+I)\neq\Gamma(-I)$, and the rectification arises. Since the main contribution to the total $\Gamma$ comes from the term corresponding to the minimum of $B^{(s)}$, we will assume below that $\Gamma\propto\exp(-B)$, where $B=\min(B^{(s)})$. When $d<d_c(I,h)$ there is only one solution to the system~\eqref{EqAlpha}, $\alpha_i^{(0)}(\eta)$. When $h\neq0$ and/or  $\gamma(x)\neq0$, we have $\alpha_1^{(0)}\neq0$, and, in contrast to the case studied in Ref.~\onlinecite{weAll}, a 2D consideration is required here.

All these properties of the bounce solutions $\alpha_i^{(s)}$ survive at finite temperatures up to some value $T^{*}$, which is the crossover temperature between the quantum and classical regimes. At low temperatures, $T<T^{*}$, the exponent $B$ only slightly decreases with $T$ (thermally-stimulated tunneling). When $T>T^{*}$, there are only imaginary-time independent solutions to Eq.~\eqref{EqAlpha}, $\alpha_i^{(s)}(\eta)=\bar{\alpha}_i^{(s)}$, where the points $(\bar{\alpha}_0^{(s)},\,\bar{\alpha}_1^{(s)})$ correspond to the extremes (minimum or saddle-points) of the potential $V(\alpha_0,\alpha_1)$ (here $s$ takes the values $s=0,\pm1$, if $d>d_c(I,h)$, and $s=0$ otherwise). Thus, when $T>T^{*}$ the exponent $B$ can be written as
\begin{equation}\label{BT}
B=\frac{9\Lambda\mu_0^{3}}{(U_{000})^2}\frac{\hbar\omega_p\left|V_r\right|}{T},\,\,\, V_{r}=\max\limits_s\left[V(\bar{\alpha}_0^{(s)},\bar{\alpha}_1^{(s)})\right].
\end{equation}
The crossover temperature $T^{*}$ is the temperature where the period of the bounce solution, $\eta_0=\sqrt{\mu_0}\hbar\omega_p/T$, becomes equal to the period of infinitesimal oscillations near the extreme point $(\bar{\alpha}_0^{(r)},\,\bar{\alpha}_1^{(r)})$, corresponding to the maximum of $V(\bar{\alpha}_0^{(s)},\bar{\alpha}_1^{(s)})$. The latter one is equal to $2\pi/\sqrt{\lambda^{+}_{r}}$, where $\lambda^{+}_{r}$ is the positive~\footnote{It can be shown that the point $(\bar{\alpha}_0^{(r)},\,\bar{\alpha}_1^{(r)})$ corresponds to the saddle-point of the potential $V(\alpha_0,\alpha_1)$, both for $d>d_c$ and $d<d_c$, so the matrix of second derivatives of $V$ at this point has only one positive eigenvalue.} eigenvalue of the matrix $\partial^2V/\partial\alpha_i\partial\alpha_j$ calculated at the point $(\bar{\alpha}_0^{(r)},\,\bar{\alpha}_1^{(r)})$. As a result, the crossover temperature can be written as $T^{*}=\hbar\omega_p\sqrt{\mu_0\lambda^{+}_{r}}/(2\pi)$.

\begin{figure}
\begin{center}
\includegraphics*[width=0.48\textwidth]{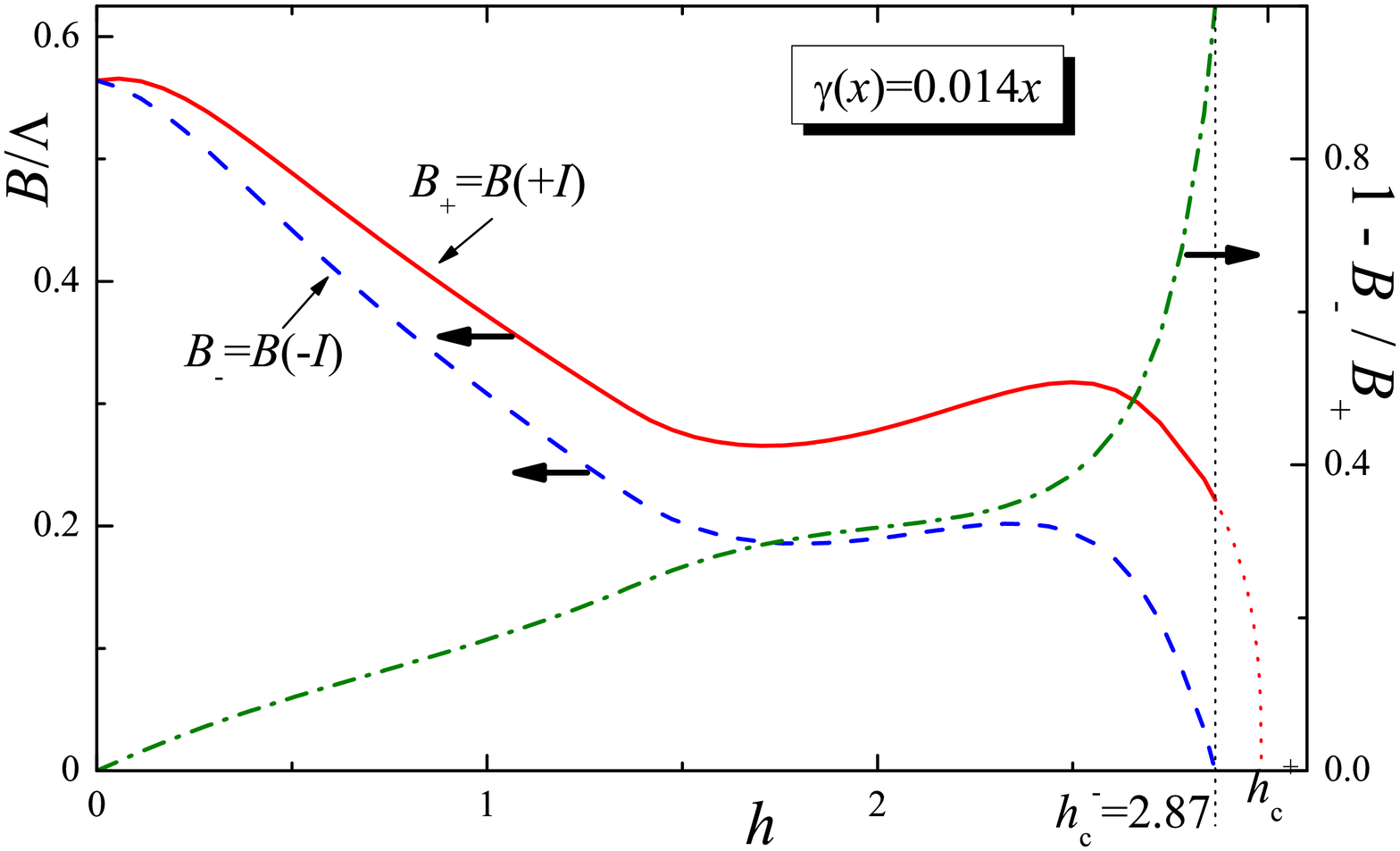}
\end{center}
\caption{\label{FigB0Hpm} (Color online) The magnetic field dependence of $B(+I,h)$ (red solid curve) and $B(-I,h)$ (blue dashed curve) for a fixed ratio $I/I_c=0.96$, calculated for $\gamma(x)=0.014x$. The relative ratio $1-B(-I,h)/B(+I,h)$ as a function of $h$ is shown by the green dotted-dashed curve. Here the parameters are $d=2.5$, $T=0$, $(\overline{\gamma^2(x)})^{1/2}\!\!\!=0.01$.}
\end{figure}

All results above were obtained for a {\it dc} current. We can also use all the above formulas for a slowly-varying current $I=I_0\cos\Omega t$, assuming that at any time $t$ the current can be considered as a dc one. This adiabatic limit is valid when the frequency $\Omega$ is much smaller than the inverse imaginary-time period, $1/\tau_0=T/\hbar$. Therefore, we obtain the condition of adiabaticity~\footnote{When $T<T^{*}$, one can use a weaker inequality, $\hbar\Omega\ll T^{*}$, since $B$ depends weakly on $T$ in this region.}: $\hbar\Omega\ll T$.

\begin{figure}\centering
   \subfigure{\includegraphics[width=0.47\textwidth]{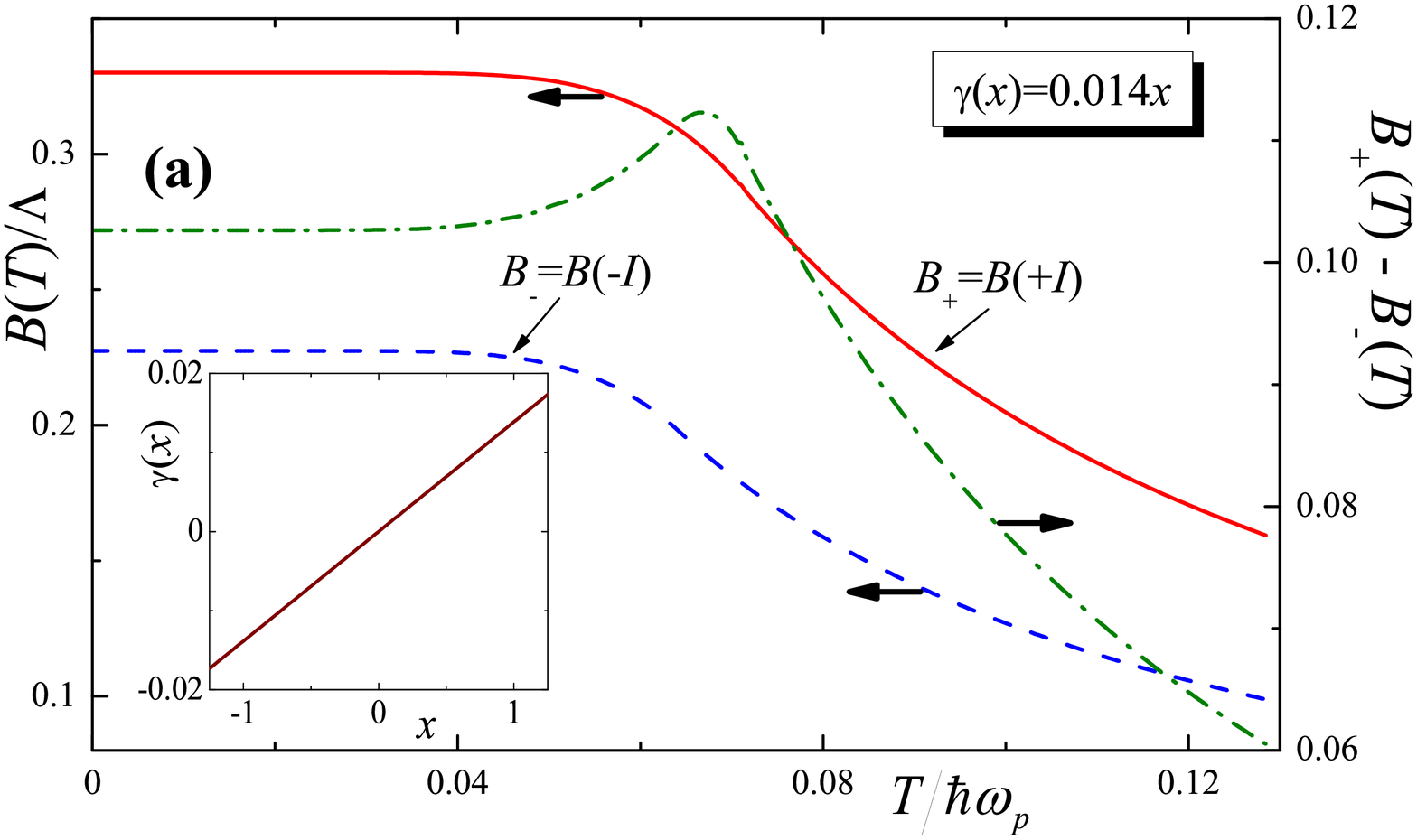}}\\
   \vspace{-4mm}\subfigure{\includegraphics[width=0.46\textwidth]{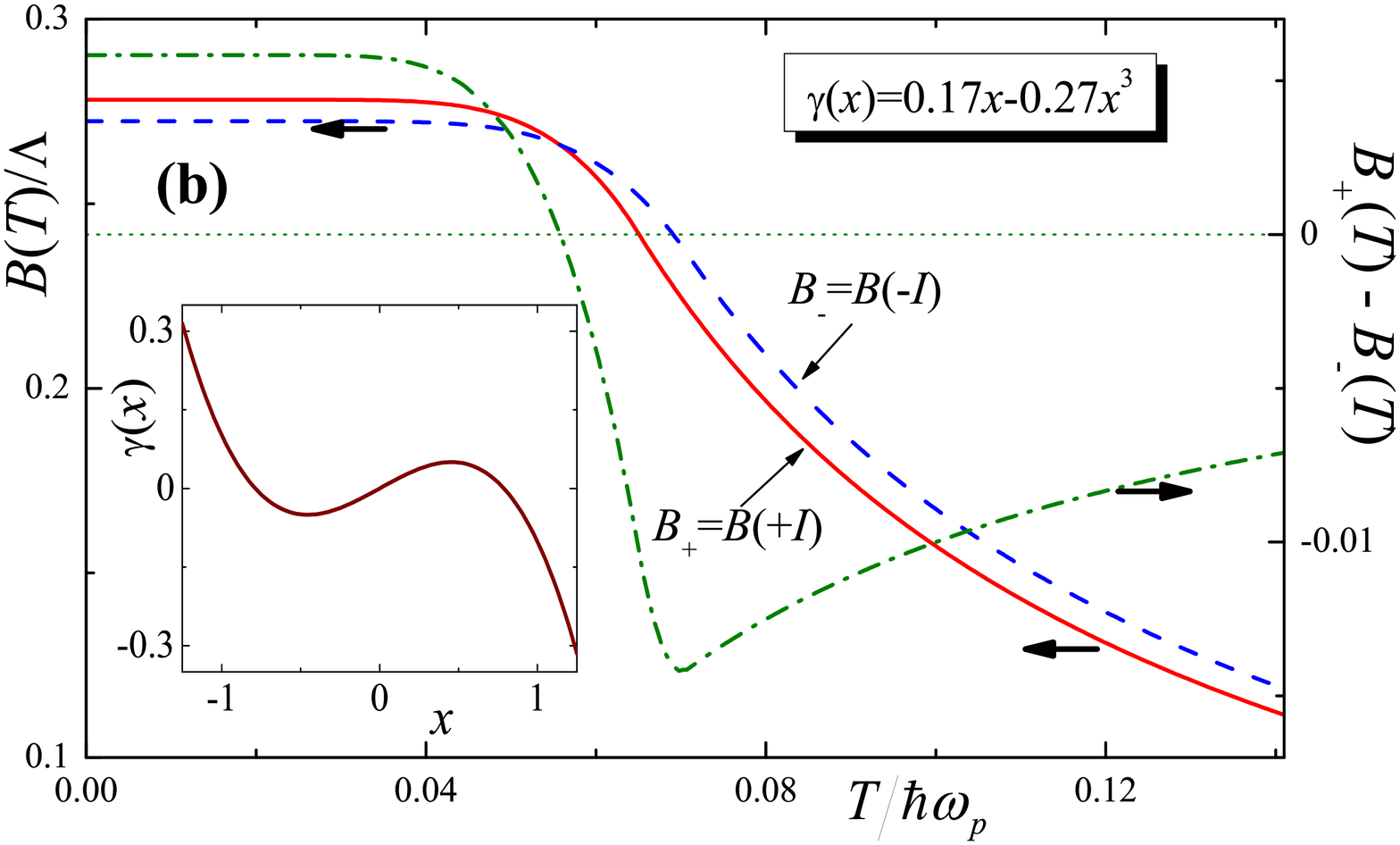}}\vspace{-3mm}
 \caption{(Color online) The escape rate exponents $B(+I)$ (red solid curves) and $B(-I)$ (blue dashed curves) versus temperature for two different spatial $\gamma(x)$ distributions (shown in the inset): $\gamma(x)=0.014x$ (a) and $\gamma(x)=0.17x-0.27x^3$ (b). The differences $B(+I)-B(-I)$ are shown by green dotted-dashed curves. The other parameters, $d=2.5$, $h=2.25$, $I/I_c=0.95$, are the same for both panels.} \label{FigBT}
\end{figure}

\textit{Results and discussions.}--- Let us first neglect the spatial distribution of the critical current density, assuming that $\gamma(x)=0$. In this case, the following condition is met $B(-I,h)=B(I,h)$, and there is no rectification. In Fig.~\ref{FigB0H}, we plot the magnetic field dependence of the exponent $B$ ($T=0$) for three fixed ratios $I/I_c(h,d)$, where the critical current $I_c(h,d)$ depends on the magnetic field. It goes to zero when $h$ achieves the critical field $h_c(d)$. Note the non-monotonic dependence of $B$ on $h$, which is related to the change in characteristic properties of the static solution $\varphi_0(x)$ with growing $h$. The non-zero $\gamma(x)$ breaks down the symmetry of $B$ with respect to the direction of the current. In this case, we have two critical currents $I_c^{\pm}(h,d)$ and two critical fields $h_c^{\pm}(d)$, corresponding to positive and negative currents. Assuming that $|\gamma(x)|\ll1$, we calculate the exponent $B$ in first order perturbation theory with respect to $\gamma$. Figure~\ref{FigB0Hpm} shows the dependence of $B(I,h)$ and $B(-I,h)$ on $h$, calculated for linear $\gamma(x)$ at the fixed ratio $I/I_c(h,d)$, where $I_c(h,d)$ is the critical current calculated to zeroth-order in $\gamma$. The degree of rectification, $1-[B(-I,h)/B(I,h)]$, shown in Fig.~\ref{FigB0Hpm}, monotonically increases up to $h=h_c^{-}$, where $I_c^{-}=0$.

Let us now consider the temperature dependence of the escape rate exponent $B$. The curves $B_{\pm}=B(\pm I)$ versus $T$, calculated for two different spatial profiles $\gamma(x)$, are shown in Fig.~\ref{FigBT}. The difference $(B_{+}-B_{-})$ which defines a ratchet effect as a function of temperature is also shown in Fig.~\ref{FigBT}. For linear $\gamma(x)$ (Fig.~\ref{FigBT}a) this difference is always positive, while for the more asymmetric profile $\gamma(x)=0.17x-0.27x^3$ (which is shown in the inset to Fig.~\ref{FigBT}b) it changes sign at temperatures near the crossover temperature between the quantum and classical regimes. There is a big difference in absolute values of the ratchet effect between these two cases: the relative ratio $(B_{+}/B_{-}-1)$ for $\gamma(x)=0.17x-0.27x^3$ (when it changes sign) is much smaller than for linear $\gamma(x)$. It can be enhanced by optimizing the profile for $\gamma(x)$. The effect of the profile $i_c(x)$ on the rectification direction can be understood following the explanation given in Ref.~\onlinecite{HanggiPRL97}: for thermoactivated hopping only the height of the potential barrier is essential, while the probability of quantum tunneling depends also on the thickness of this barrier.

The sign of the difference $(B_{+}-B_{-})$ defines the sign of the rectified voltage $V_{dc}$. Thus, the effect of the change of sign of $(B_{+}-B_{-})$ can be observed by measuring the dc voltage as a function of temperature. The value $(B_{+}-B_{-})$ can be calculated using data of the escape rate $\Gamma$ obtained either in ac or dc current measurements.

\textit{Conclusions.}--- We have proposed a new type of ratchet system based on a long Josephson junction with a spatially inhomogeneous critical current density $i_c(x)$. The exponent of the escape rate of the phase difference $\varphi$ was calculated as a function of: temperature, dc magnetic field, and dc or slow ac current. We have shown that due to both the magnetic field and the spatial inhomogeneity of $i_c$, the escape rate becomes asymmetric with respect to the direction of the current. This leads, in particular, to the appearance of a dc voltage when the system is biased by an ac current. We have also shown that, for definite shapes of $i_c(x)$, the rectified voltage changes sign at a temperature near the crossover temperature between the quantum and classical regimes.

This work was supported in part by the RFBR (projects JSPS-RFBR 09-02-92114 and 09-02-00248), the US NSA, LPS, ARO, and NSF. SS acknowledges support from the EPSRC via EP/D072581/1 and EP/F00548211. AOS acknowledges support from the Russian Science Support Foundation.

\end{document}